# From PDFs to Structured Data: Utilizing LLM Analysis in Sports Database Management

Juhani Merilehto

Jyväskylä University of Applied Sciences, School of Social and Health Studies

Likes Institute

Jyväskylä, Finland

{juhani.merilehto}@jamk.fi

**Abstract.** This study investigates the effectiveness of Large Language Models (LLMs) in processing semi-structured data from PDF documents into structured formats, specifically examining their application in updating the Finnish Sports Clubs Database. Through action research methodology, we developed and evaluated an AI-assisted approach utilizing OpenAI's GPT-4 and Anthropic's Claude 3 Opus models to process data from 72 sports federation membership reports. The system achieved a 90% success rate in automated processing, successfully handling 65 of 72 files without errors and converting over 7,900 rows of data. While the initial development time was comparable to traditional manual processing (three months), the implemented system shows potential for reducing future processing time by approximately 90%. Key challenges included handling multilingual content, processing multi-page datasets, and managing extraneous information. The findings suggest that while LLMs demonstrate significant potential for automating semi-structured data processing tasks, optimal results are achieved through a hybrid approach combining AI automation with selective human oversight. This research contributes to the growing body of literature on practical LLM applications in organizational data management and provides insights into the transformation of traditional data processing workflows.

## 1  Introduction

The advancement of Generative Artificial Intelligence (Jovanovic & Campbell, 2022), most notably via the Large Language Models (Zhou et al. 2023), has opened new avenues for data analysis across various fields (e.g., Shang & Huang, 2024; Linkon et al. 2024; Salah et al. 2023; DuPre & Poldrack, 2024).

The rapid advancement of Artificial Intelligence (AI), particularly in the domain of Generative AI (Jovanovic & Campbell, 2022), has created a new era of data analysis capabilities. At the forefront of this revolution are Large Language Models (LLMs), which have demonstrated remarkable proficiency in understanding and generating human-like text across diverse domains (Zhou et al., 2023). These models, trained on large amounts of text data, have shown potential to be utilized in various fields of research and practice, from business management (Linkon et al., 2024) to social psychology (Salah et al., 2023) and neuroimaging (DuPre & Poldrack, 2024).

In the domain of data analysis, LLMs offer capabilities that extend beyond traditional statistical methods. Their ability to interpret context, understand natural language, and generate coherent responses makes them particularly suited for tasks involving unstructured or semi-structured data (Katz et al. 2024). This potential for data-analysis has been recognized across various disciplines (Inala et al. 2024), with researchers exploring LLMs' applications in areas such as generative graph analytics (Shang & Huang, 2024).

This study continues the line of inquiry for practical use-cases for LLMs in real life data-analysis scenarios. Specifically, this study focuses on the challenge of extracting and structuring information from PDF reports which include structured and semi-structured information as lists or non-ideally formatted tables. The handling (i.e., compiling) and analyzing such data usually requires cleaning and gathering them to other formats (i.e., excel-files or other structured tabular formats). This task, traditionally labor-intensive and prone to human error, serves as an ideal test case for the capabilities of LLMs in real-world data processing scenarios.

To this end, two foundation model LLMs (in 2024) were utilized: OpenAI's GPT-4 and Anthropic's Claude 3 Opus. These models were tasked with analyzing and structuring data from PDF reports, a common format for organizational reporting that often presents challenges in data extraction and analysis. By comparing the performance of these LLMs against traditional manual data processing methods that were conducted in the previous year on a similar task, this study aims to assess the potential of AI-driven approaches in enhancing the efficiency and accuracy of data-analysis with semi-structured PDF data.

This research contributes to the growing body of literature on practical applications of LLMs but also offers insights into the potential transformation of data analysis work in organizations.

## 2  Related Work

The use of Large Language Models (LLMs) in data analysis has gained significant traction in recent years, marking a paradigm shift in how we approach complex data processing tasks. This section reviews key developments in the application of LLMs to data analysis, focusing on their capabilities, limitations, and the evolving methodologies for their effective utilization.

For example, in the work of Brown et al. (2020) discussing GPT-3, it was demonstrated that capabilities in natural language processing tasks, including data interpretation and generation had taken significant steps forward. This work laid one of the foundations for exploring LLMs in various data analysis contexts, showcasing their potential to understand and generate human-like text based on vast amounts of training data. In the extensive work of Bommasani et al. (2021), they provided a comprehensive overview of the potential applications and ethical considerations of large language models in general use, terming them "foundation models." Their work highlighted the transformative potential of these models across various domains, including data analysis, while also emphasizing the need for careful consideration of their limitations and societal impacts.

Addressing the challenge of domain-specific knowledge in data analysis, Wei et al. (2022) introduced the concept of "chain-of-thought prompting" for LLMs. This technique showed promise in enhancing the reasoning capabilities of LLMs when dealing with complex, multi-step data analysis tasks, potentially making them more reliable tools for specialized analytical work.

However, the application of LLMs in data analysis in general is not without challenges. E.g., Bender et al. (2021) raised important questions about the limitations and potential risks of relying on large language models, particularly in contexts where accuracy and interpretability are crucial. Their work (2022) underscores the importance of developing robust methodologies for validating and interpreting LLM outputs in data analysis contexts.

Recently LLMs have been used for a variety data analysis tasks (Inala et al. 2024) that have previously been more in the domain of human experts, i.e. in Thematic Analysis (Katz et al., 2024). LLMs and Generative AI is not however, a solely binary decision of dividing work between a human and the machine, but is more than often a collaborative effort (Drosos et al. 2024). Collaboration in work tasks has been seen to increase performance in organizational work context remarkably (see i.e., Brynjolfsson et al. 2023 and Dell'Acqua et al. 2023).

The literature reveals an evolving landscape where, among other use-cases, LLMs are increasingly being integrated into data analysis workflows. While these models offer powerful capabilities for processing and interpreting complex data, researchers consistently emphasize the need for careful validation, and domain-specific challenges (Gu et al. 2024).

## 3  Methodology and Procedure

This study employed an action research methodology to develop and refine an AI-assisted approach for converting semi-structured PDF data from Finnish sports unions into structured Excel files. Action research was chosen for its suitability in addressing practical problems while simultaneously generating new knowledge about the process (Reason & Bradbury, 2008). The research followed a cyclical process characteristic of action research, comprising planning, action, observation, and reflection stages (Kemmis & McTaggart, 2008). This iterative approach allowed for continuous improvement of the data extraction and structuring process.

As both the researcher and the practitioner, I adopted a participant-observer role. This dual position enabled direct engagement with the technical challenges while maintaining a critical, analytical perspective on the process and outcomes (Coghlan & Brannick, 2014). Data sources included PDF documents from 72 Finnish sports unions containing semi-structured information about member clubs, output from the AI-assisted data extraction process, and detailed notes on the performance and issues encountered for each union's data processing. Data collection involved processing each union's PDF document through the developed system and recording both quantitative metrics (e.g., accuracy rates) and qualitative observations (e.g., specific data handling challenges).

The technical solution[1] leveraged first the OpenAI API and the Anthropic API. Specifically, the Anthropic models was the Claude-3-opus. Key components of the implementation also included PDF text extraction using PyMuPDF, API interaction using a carefully crafted and iteratively tested prompt, response parsing and structuring using pandas, and output generation in Excel format (see Figure 1). While technically the prompt was zero-shot (Kojima et al., 2022), the instructions included some examples of how to interpret different articulations of headings in tables, i.e., what to prioritize. The system was developed iteratively, with each cycle informing refinements to the code, API prompts, and overall process.

The development followed a cyclical process of planning (identifying areas for improvement based on previous results), action (implementing changes to the code or prompts), observation (processing a set of union PDFs and collecting performance data), and reflection (analyzing results, identifying successes and challenges). This cycle was repeated multiple times, with each iteration documented to track the evolution of the solution. The entire development and analysis process was possible to compare to the previous year's process, which was done mostly manually.

---

[1] GitHub repository: juhanimerilehto. (15.10.2024). simple-pdf-analysis. GitHub. https://github.com/juhanimerilehto/simple-pdf-analysis

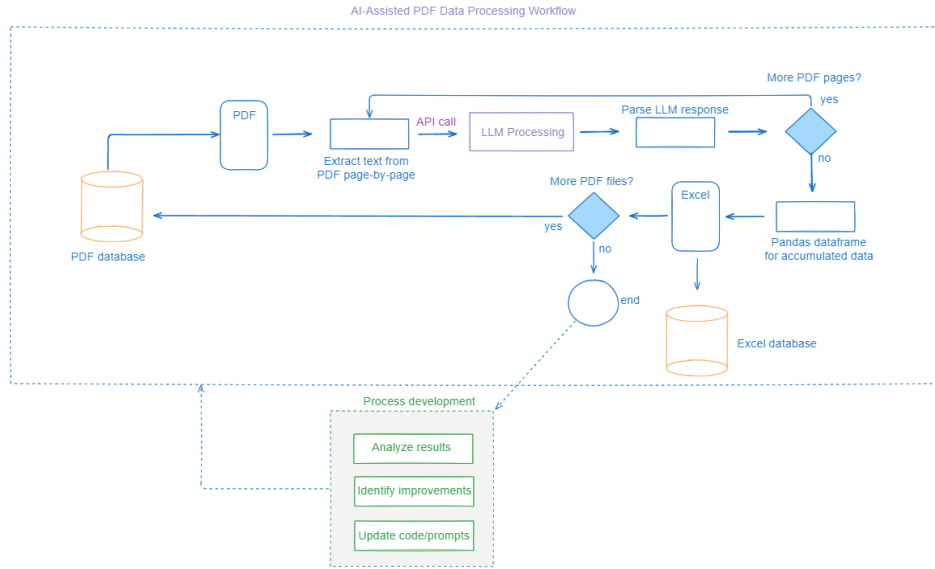

**Figure 1: AI-Assisted PDF Data Processing and Development**

The effectiveness of the AI-assisted approach was evaluated using several criteria: accuracy (percentage of correctly extracted and structured data points), completeness (ability to handle all required information such as club name, Business ID, and member count), and robustness (performance across varied PDF formats and data presentations).

By employing this action research methodology, the study aimed to not only develop a practical solution for data structuring but also to generate insights into the process of integrating AI technologies in data management tasks within specific organizational contexts. This approach allowed for a deep understanding of the challenges and opportunities presented by AI-assisted data processing.

Ethical considerations included ensuring data privacy and confidentiality, as well as maintaining transparency about the AI-assisted nature of the data processing. No personal data was processed, and all information was treated as organizational-level data. The study acknowledges potential limitations, including researcher bias due to the dual role of developer and analyst, generalizability constraints given the focus on Finnish sports union data, and dependence on the capabilities and potential biases of the chosen AI model. These limitations were mitigated through rigorous documentation, regular self-reflection, and transparency in reporting both successes and challenges.

# 4  Process and Results

The underlying task was to update the Finnish Sports Clubs Database (Seuratietokanta, 2024), upkept by Likes Institute in the Jyväskylä School of Applied Sciences, Finland (Likes, 2024). The task is annual and was first created in 2023. The sports club database is a public database intended for anyone interested in sports and exercise club activities. It includes member clubs from sports federations that received general funding from the Finnish Ministry of Education and Culture in 2023 and 2024 respectively. Currently, the details of a total of 8,188 sports and exercise clubs from 74 different sports unions are listed.

The Large Language Model (LLM)-based approach for converting semi-structured PDF data from sports unions into fully structured Excel files demonstrated promising results across the 72 analyzed unions (two unions were excluded from this analysis). The system exhibited high accuracy for many unions, with several achieving 100% correct data conversion. However, the analysis revealed some consistent challenges and areas for improvement.

One of the most prevalent issues was the "double name problem," occurring when entities had names in multiple languages (typically Finnish and Swedish) written in the same cell or included additional information in the same cell such as location. This often resulted in data misalignment, pushing other crucial information like business IDs (Y-tunnus) or member counts into incorrect columns. Another recurring challenge was the inclusion of extraneous information, such as headers (e.g., "OKM", acronym for the Finnish Ministry of Education) or summary rows (aka., "totals"), as data entries.

The system showed adaptability in handling various data formats, successfully processing both complex tabular structures and list formats. However, larger datasets (i.e., over 400 rows), presented difficulties when processing multiple pages simultaneously. The errors encountered were mostly missing information (i.e., 10 % results missing) and hallucinations, while rare, on club's names and member counts. This could be handled by feeding the pages to the API LLM one by one, practically removing the issue entirely.

Iterative improvements to the prompt, including addressing missing information and removing unnecessary "introductory" text, led to enhanced performance. The prompt was also tuned to pick out the correct information out of several potential ones – for example, a sports club could have listed different types of members, some of them could be a count of honorary members, or subsect of the members (i.e., those with sports-related licenses). While most unions were processed with high accuracy, some unique cases as previously mentioned emerged.

The first model utilized was OpenAI ChatGPT 4 (gpt-4-0613). The usage of the Python script and API was a straightforward process, with little technical challenges. The context-range at the time of testing was 128k tokens, with a maximum response size of 4k tokens, which were more than enough for the use-cases for this project. After the first iterations, the solution was changed to utilize the Anthropic's Claude 3 Opus – model, which proved to be roughly similar in capabilities and technical usability. The context windows were roughly similar, input being allocated at 200k tokens and output at 4k tokens. Higher-capability state-of-the-art models were utilized since there were no needs for considering the speed of the response or the token cost of queries.

Overall, the LLM-based approach proved effective for structuring sports union data, with many unions achieving near-perfect or perfect accuracy. During the iterative process, the first successful batch had approximately half of the PDF's correctly formatted (37 files, total of over 3000 rows), but by consecutive improvements the final iteration proved to be approximately 90 % successful in its processing. This meant that 65 of 72 files were processed without errors. Of the remaining 7 files, two had to be input entirely manually and 5 corrected manually. The result was over 7,900 rows of data correctly processed by the LLM-pipeline.

However, there remain opportunities for refinement, particularly in handling multi-language names, ensuring consistent column alignment across varied table structures, filtering non-data rows, and processing extensive multi-page datasets. These insights provide a clear direction for future improvements to enhance the system's accuracy and reliability in converting semi-structured PDF data to structured Excel formats.

## 5 Discussion

This study explored the application of an AI-assisted approach to convert semi-structured PDF data from Finnish sports unions into structured Excel files, employing action research methodology. The process involved developing a Python script that leveraged the Anthropic API and the Claude-3-opus model, requiring a preliminary level of technical skills and an understanding of Large Language Model (LLM) capabilities and functioning. This approach marks a significant shift from the previous year's predominantly manual data input method, which relied on Adobe Acrobat's table conversion function for well-structured tables and manual typing of data from PDF to Excel, followed by manual verification.

*Efficiency and Time Investment*

In total there was a similarity in overall time investment between the AI-assisted approach and the previous manual method, with both taking approximately three months. However, the nature of this time investment differed significantly. While the manual approach required sustained data entry and verification efforts, the AI-assisted method's time was largely devoted to the iterative development of the LLM pipeline and rigorous testing. This distinction is crucial, as it highlights a shift from repetitive manual labor to a more front-loaded, skill-intensive process of system development.

Importantly, once the AI-assisted system was finalized and optimized, the potential for future efficiency gains became apparent. The study suggests that with the developed system in place, subsequent data processing tasks of similar scale could be completed in approximately one-tenth of the time taken for the entire project. This dramatic reduction in processing time represents a significant potential for increased efficiency in sports organization data management.

*Technical Skills and Interdisciplinary Knowledge*

The development of the AI-assisted approach required a blend of technical skills and domain-specific knowledge. The need for Python programming skills, API integration experience, and an understanding of LLM capabilities highlights the increasingly interdisciplinary nature of data management in sports organizations. This shift towards more technical solutions may necessitate upskilling or new hiring strategies within sports management bodies to fully leverage AI and automation technologies.

*Challenges and Adaptations*

The iterative nature of the action research methodology allowed for continuous refinement of the AI-assisted approach. Challenges encountered during the process, such as handling varied PDF formats and inconsistent data presentations across different sports unions, necessitated ongoing adjustments to the Python script and API prompts. This adaptability was key to developing a robust solution capable of handling the diverse data structures present in the sports union documents.

*Comparative Analysis with Manual Methods*

While the overall time investment was similar, the AI-assisted approach offered several advantages over the manual method. The automation of data extraction and structuring reduced the potential for human error in repetitive tasks. Additionally, the consistent application of predefined rules through the API prompts ensured a more standardized approach to data interpretation across all documents, potentially improving data consistency and reliability.

However, the study also revealed that certain well-structured tables could be efficiently converted using an LLM, there yet remained some data which were too unstructured and thus required some manual approach. This finding suggests that a hybrid approach (e.g., Drosos et al. 2024), combining AI-assisted processing with selective use manual assistance, might offer an optimal balance of efficiency and accuracy – at least until the next generation of LLM's are capable of handling it fully.

*Limitations and Future Research*

Despite its promising results, this study has limitations that warrant consideration. The focus on Finnish sports union data may limit the generalizability of the findings to other contexts or data types. Additionally, the reliance on a specific AI model (Claude-3-opus) raises questions about the approach's adaptability to other AI technologies or future model updates.

Future research could explore more in-depth the application of this AI-assisted approach to a broader range of sports management data, including extensive diary of different prompts, data examples, and rates of successes when using them. While this study scraped the surface of demonstrating the potential of AI-assisted approaches in transforming data management practices, it nevertheless brought additional research into the LLM data analysis literature.

# 6 Conclusion

This study demonstrates the potential of Large Language Models in transforming semi-structured data processing tasks. Through systematic implementation of an LLM-based approach using Claude 3 Opus and GPT-4 models, it was recognized that AI-assisted data processing achieved near-perfect accuracy in data structuring for semi-structured data analysis. While the initial time investment was comparable to manual processing (approximately three months), the developed system shows promise for dramatic efficiency improvements in future iterations, potentially reducing processing time by 90% for similar tasks.

However, several challenges emerged during implementation, particularly in handling multilingual content (Finnish-Swedish combinations), processing extensive multi-page datasets, and managing extraneous information. These limitations suggest that, at least in the current state of LLM technology, a hybrid approach combining AI-assisted processing with selective human oversight may be optimal for ensuring data quality and reliability. The successful implementation of such systems requires both technical expertise and domain knowledge, indicating that organizations may need to invest in upskilling or new hiring strategies to fully leverage these technologies.

Looking forward, this research opens several avenues for future investigation while demonstrating the practical viability of LLM-based approaches for organizational data processing tasks. As LLM technology continues to evolve, organizations that develop the necessary technical capabilities and understanding to leverage these tools effectively will be well-positioned to realize significant improvements in their data processing workflows. The findings suggest that while LLMs are not yet a complete replacement for human oversight in data processing, they represent a powerful tool for enhancing efficiency and accuracy in organizational data management.

**Acknowledgements**

This research was conducted at Likes Institute, the Research Centre for Physical Activity and Health, part of JAMK University of Applied Sciences.

**Funding**: This work was conducted as part of the author's responsibilities at Likes Institute, Jamk.

**Data Availability:** The Finnish Sports Clubs Database (Seuratietokanta) is publicly available through Likes Institute, and a version of the technical solutions is openly available at https://github.com/juhanimerilehto/simple-pdf-analysis

**Conflicts of Interest:** The author declares no conflict of interest.